\documentclass[a4paper, notoc]{JHEP3}
\usepackage[dvips]{graphicx}
\usepackage{amsfonts}
\usepackage{amssymb}
\usepackage{epsfig}
\usepackage{cite}

\newcommand{\be}{\begin{equation}}
\newcommand{\ee}{\end{equation}}
\newcommand{\bea}{\begin{eqnarray}}
\newcommand{\eea}{\end{eqnarray}}
\newcommand{\nn}{\nonumber}

\newcommand{\ti}{\times}
\newcommand{\half}{\frac{1}{2}}
\newcommand{\mc}{\mathcal}

\newcommand{\beqa}{\begin{eqnarray}}
\newcommand{\eeqa}{\end{eqnarray}}

 \title{Quantum Gravity Constraints on Inflation}
\author{Joseph P. Conlon
 \\ Rudolf Peierls Centre for Theoretical Physics, 1 Keble Road \\
 Oxford OX1 3NP, UK \\ Email: \email{j.conlon1@physics.ox.ac.uk}}

\abstract{We study quantum gravity constraints on inflationary model building. Our approach is
based on requiring the entropy associated to a given inflationary model to be less than that
of the de Sitter entropy. We give two prescriptions for determining the inflationary entropy, based on
either `bits per unit area' or entanglement entropy.
The existence of transPlanckian flat directions, necessary for large tensor modes in the CMB,
correlates with an inflationary entropy greater than that allowed by de Sitter space.
Independently these techniques also constrain or exclude de Sitter models with large-rank gauge groups and high
UV cutoffs, such as racetrack inflation or the KKLT construction.
}

\preprint{}

\begin{document}

\section{Introduction and Review}

Inflation is a powerful and compelling explanation for both the large-scale homogeneity and flatness of the universe and also for the origin
of the small density perturbations that lead to the growth of structure. This is a wonderful story,
 the physics of which is reviewed in \cite{Mukhanov, 0503203, LiddleLyth, 09075424}.
Among physical mechanisms with some degree
of observational support, inflation also probes the largest energy scales: the inflationary energy scale in many models
is $V_{inf} \sim (10^{13} \to 10^{15} \hbox{GeV})^4$.
However although inflation is a
quantum phenomenon that involves gravity, the conventional
formulation is only quantum field-theoretic. Density perturbations are generated by quantum fluctuations within the approximate
Minkowski space deep inside the horizon, and grow through the classical de Sitter expansion. Given the high energies and presence of gravity,
it is natural to consider restrictions on inflation from quantum gravity or Planck scale physics, for example string theory.

The most simple-minded way to search for constraints is by trying to build consistent inflationary models in string theory and seeing what is found.
Reviews of these models and their constructions can be found in \cite{0702059, 09010265, 11082659, 11082660}.
The iteration of such attempts has indicated some possible constraints. Our focus here is
  the possibility that string theory may forbid models with large tensors.\footnote{However 
  for some alternative proposals to generate tensor modes see \cite{09072384, 11105389}.}
  
This is equivalent to the statement that string theory does not admit
inflationary potentials which are flat over parametrically trans-Planckian field ranges.

There are two elements to this statement. The top-down element is that field theory modes arising from concrete string compactifications
are often limited to sub-Planckian ranges. Examples are D3 brane position moduli \cite{0610285} or the size of internal cycles in a Calabi-Yau
manifold. A transPlanckian D3 brane position excursion corresponds to movement through a distance greater than the Calabi-Yau radius.
Moduli which can have transPlanckian displacements - for example the volume and the dilaton - couple to all modes of the theory.
Moving such string moduli through
 transPlanckian distances causes large changes in the spectrum, the appearance of new light modes and
  the breakdown of the original effective field theory. Furthermore any stringy proposals for generating
  transPlanckian vevs run into control issues. Examples are
 N-flation \cite{0507205} or axion monodromy inflation \cite{08033085, 08080706}. The former involves the
 combination of a large number of axions with sub-Planckian field range
 to generate a trans-Planckian field range. This model is vulnerable to corrections proportional to the number of species.
 Another proposal is axion monodromy inflation which
 suffers from a large backreaction from the 5-brane/anti-5-brane interaction \cite{11106454}.

The bottom-up element is that
in field theory one can simply write down potentials with parametrically trans-Planckian flatness.
However these examples - such as natural inflation
with superPlanckian axion decay constants - have properties not realised in string theory.
Other field theory models such as chaotic inflation suffer in string theory from an inability to control the coefficients
 of Planck-suppressed corrections $\sum_n \lambda_n \left( \frac{\phi}{M_P} \right)^n$.

String theory therefore appears to censor attempts to realise large field inflation. As string theory is a theory of quantum gravity,
it is natural to wonder whether this censorship may be quantum gravitational in nature, with the repeated
failure to find large field inflation within string theory
a reflection of deeper underlying principles rather than a failure to try hard enough.
This is not unreasonable.
Inflation involves (at least to a good approximation) de Sitter space, and de Sitter space is a quantum gravitational object in the same
way that a black hole is. The basic physics of inflation is that of flat directions in de Sitter space.
The purpose of this paper to investigate quantum gravitational bounds on such flat directions,
and in particular on the allowed number and range of such directions.

Our approach is entropic in nature. As a quantum gravity object,
de Sitter space has a finite horizon, a finite entropy, and therefore a finite number of microstates.
However we can also associate an entropy to the light fields necessary for inflation,
which heuristically should grow with both the number of fields and the allowed
field range. The thrust of this paper is that for sufficiently many fields and sufficiently large field range, these fields give a calculable entropy
in excess of the de Sitter entropy. For some earlier related work, see \cite{0211221, 0301182, 0306070, cites}.

We will find that this approach can be used more generally. It also constrains subPlanckian inflationary models with high cutoff scales and
large numbers of degrees of freedom (for example racetrack inflation \cite{0406230}). Finally it also bounds pure
constructions of de Sitter vacua with high string scales and relying on the condensation of large-rank gauge groups (such as the
KKLT scenario \cite{0301240}).

The paper is structured as follows. In section 2 we briefly discuss de Sitter space in string theory and frame our problem. In section 3
we discuss de Sitter entropy and give two prescriptions for computing the entropy, one based on `bits per unit area' and one on
entanglement entropy. In section 4 and 5 we explain how to determine the ultraviolet cutoff and how this affects models attempting
to realise transPlanckian field vevs. In section 6 we study constraints on de Sitter vacua and more general inflationary models, before
concluding.

\section{de Sitter Space in String Theory}

There exist a variety of proposed string theory constructions of de Sitter space of varied credibility.
These constructions are reviewed in \cite{0405068, 0610102, 0611039, 08031194, 09070665} and details of some original constructions can be found in
\cite{0301240, 0408054, 0502058, 0505076, 0701034, 07121196, 10031982}.
A common feature is the presence of many axions, which due to a perturbative continuous shift symmetry remain
essentially flat even after moduli stabilisation, with nonperturbative effects smaller than
any relevant scale. For example, the LARGE volume models \cite{0502058, 0505076}, characterised by a volume $\mc{V} \gg 1$, always contain an axion associated with the volume modulus. Any nonperturbative effects lifting this axion are suppressed by a scale $M_P e^{-2 \pi \mc{V}^{2/3}}$.
The exponentially large size of $\mc{V}$ implies this exponent can easily be suppressed by a factor $e^{-10^{10}}$ which
is entirely negligible.

The upshot is that if constructions of de Sitter vacua in string theory are correct,
there exist de Sitter vacua in quantum gravity which semiclassically come in continuous families parametrised by
axionic coordinates of finite field range. In the string theory constructions
both the number of axions and their field range vary between models. In this case quantum gravity
has states that are at low energy semiclassically described as
\be
\label{qg}
\mc{L} = \frac{M_P^2}{2} \int d^4x \sqrt{g} \left( R - V - \half \sum_{a=1}^N \partial_{\mu} \phi_a \partial^{\mu} \phi_a \right),
\qquad \textrm{ with $\phi_a \equiv \phi_a + f_a$.}
\ee
Classically there are an infinite number of such states, the infinite number being associated to the vev of the axions $\phi_a$.

However in quantum gravity de Sitter space has a finite entropy \cite{GibbonsHawking77}
$$
S_{dS} = 8 \pi^2 \frac{M_P^2}{H^2}.
$$
Here $M_P$ is the reduced Planck mass.
This entropy implies a finite number of quantum microstates, and so in quantum gravity the
semi-classical description of (\ref{qg}) must break down to leave a finite number of microstates.

The basic problem can now be formulated without any reference to string theory. Consider de Sitter space,
at a Hubble scale $H \ll M_P$, minimally coupled to a set of $N$ massless
axion fields $\phi_a, a = 1 \ldots N$ of field range $f_a$ with $H \ll f_a \ll M_P$.
As both $f_a, H \ll M_P$ we expect
semiclassical approximations to work well.
We must also be able to associate some number of
de Sitter microstates $\mc{N}$ to the axion fields. How does $\mc{N}$ depend on $f_a$ and $N$?

Intuitively we expect that the more fields present, the more potential microstates there are (so $S$ will increase with $N$). We also expect
the number of microstates to increase as $\phi_a$ can take on more values (so $S$ will increase with $f_a$).
The requirement that the total number of
 microstates be smaller than that associated to the de Sitter entropy will then provide a bound on values of $N$ and $f_a$
 consistent with quantum gravity.

We will argue that $N$ fields of field range $f_a$ give a contribution to the entropy of
$$
S = \lambda \sum_a \frac{f_a^2}{H^2},
$$
where $\lambda$ is an $\mc{O}(1)$ number whose precise value depends on how we count microstates.
Requiring that $S$ is smaller than the de Sitter entropy then gives a parametric bound
$$
\sum_a f_a^2 < \mc{O}(1) M_P^2.
$$
In particular, this constrains models such as N-flation where N axionic directions are combined to create an effective transPlanckian field range $\sqrt{N} f_a > M_P$.\footnote{Constraints from number of species are discussed in \cite{0507205}, but in terms of a renormalisation
of Newton's constant that is model-dependent and which for specific models could be very small due to cancellations.}

We make one brief aside here. Once you allow two axion fields large field excursions can formally
be constructed by multi-wrappings on the $(\phi_1, \phi_2)$
torus, e.g. by considering an axion moving on the trajectory from $(0,0)$ to $(p f_a, q f_a)$, where $p$ and $q$ are relatively prime.
Such trajectories appear artificial - for large
$p$ and $q$ such trajectories become densely packed and recur close to points previously passed. It is also difficult to see how to
construct reasonable potentials such that this trajectory could represent the minimum. For these reasons, while we do not have a rigorous argument
to exclude such trajectories we shall not consider them further.

\section{Counting Microstates}

The basic meaning of the entropy of a system is the logarithm of the number of microstates associated to that system.
Suppose we have a construction of de Sitter space in string theory. Over sufficient time,
the de Sitter expansion will exponentially dilute any initial matter content and the far future of the system consists of an
asymptotic quantum object `de Sitter space', independent of the details of the initial field configurations.
This object `de Sitter space' may have some number of massless or light fields which will have a thermal spectrum set by the
de Sitter temperature.

We will take as our working microphysical definition of the entropy of de Sitter space
the number of initial states that can be prepared that will evolve in the far future to this quantum mechanical object.
That is, suppose we have a set of $N$ fields $\phi_1(x), \phi_2(x), \ldots \phi_N(x)$ with masses below the ultraviolet cutoff.
At initial time $t=0$ there are many possible choices of profile for these fields, with energy sourced in the field gradients. Over time
the de Sitter expansion will dilute the gradient energy and the field profile at future infinity is set by thermal fluctuations about the vacuum.
For masses greater than the Hubble scale $m \gtrsim T_{ds} = \frac{H}{2 \pi}$, the field thermally bumps around the vacuum. For massless fields
of finite field range $f_a$, the thermal fluctuations cause an eventual delocalisation of the wavefunction vev around the vacuum manifold of the field. During each efolding the fluctuations in the vev are $\delta \phi \sim H$, and so the field vev undergoes a random walk with
$\langle (\delta \phi)^2 \rangle \sim N H^2$ after $N$ e-foldings. As we evolve into the far future $N \to \infty$, and
once $N H^2 \gtrsim f_a^2$ the field vev becomes entirely delocalised. We take the entropy to be
given by the total number of initial field configurations
consistent with this future de Sitter state.

 When constructing the initial field configuration we are classically free to turn on small gradients in $\phi_a(x)$ or modify $\langle \phi_a(x) \rangle$. Provided the energy associated to these configurations is small, the de Sitter expansion will dilute this energy and the asymptotic configuration will be unaffected. Classically there are an infinite number of initial field profiles leading to an infinite de Sitter entropy. However in quantum mechanics the spectrum of modes is quantised leading to a finite entropy.

It is crucial that the future evolution leads to the asymptotic de Sitter state. If this is not the case, then the initial configuration cannot
be counted as a de Sitter microstate and cannot contribute to the entropy. The requirement that future evolution leads to de Sitter places an upper bound on the energy that can be placed in the initial axion field configuration. In string models, de Sitter space is always metastable: there is always a barrier between de Sitter space and the 10-dimensional Minkowski space solution. This is illustrated in figure 1.
\begin{figure}
\begin{center}
\epsfig{file=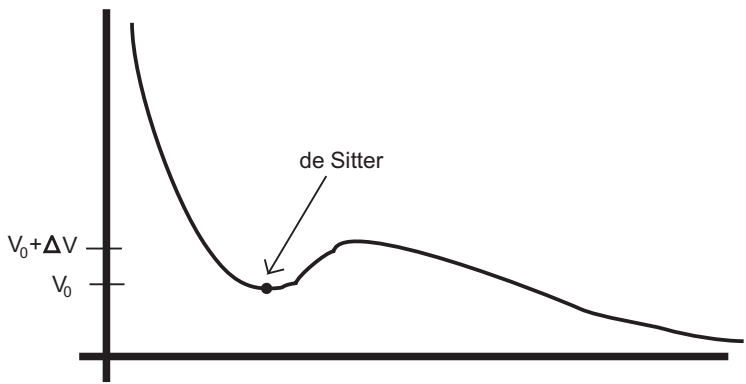,height=5cm}
\caption{The realisation of de Sitter space in string theory: de Sitter space exists as a metastable object with a barrier between
it and the 10-dimensional decompactification solution.}
\end{center}
\end{figure}
There are two scales present: the potential $V$ and the barrier height $\Delta V$. The `generic' case is that $\Delta V \sim V$, with a barrier
height comparable to the Hubble scale: to do otherwise requires fine-tuning. However one can also imagine cases where $\Delta V \gg V$ (this is very
probably the case in our universe) or $\Delta V \ll V$ (i.e. a very shallow barrier). The minimum effective barrier is $\Delta V > H^4$; if this is not satisfied than thermal effects will be sufficient to carry $\phi$ over the barrier and de Sitter space does not exist as a
long-lived metastable solution.
In string theory this decompactification barrier is associated to the volume or dilaton fields, and all sources of energy couple to these
fields.\footnote{As a simple example, in IIB the K\"ahler potential of the volume modulus is $K = - 3 \ln(T + \bar{T})$, and so the axion
kinetic term is $\int d^4 x \sqrt{g} \frac{\partial_{\mu} \theta \partial^{\mu} \theta}{4 \tau^2}$, for $\tau = \hbox{Re}(T)$. A gradient
profile for $(\partial_{\mu} \theta)^2$ therefore sources the potential for $\tau$ and sufficiently large gradients
will destabilise it.}

To avoid destabilisation and to ensure that the initial configuration asymptotes to de Sitter space, we require that the initial energy density in the field configuration is less than that of the barrier:
\be
f_a^2 \partial_{\mu} \theta \partial^{\mu} \theta \lesssim \Delta V.
\ee
We do not at this stage specify $\Delta V$.

There are (at least) two distinguishable contributions to the entropy. The entropy counts all field configurations that asymptote to future
de Sitter. We can separate these into field configurations that are localised in the interior, and so do not affect the field profile near
the de Sitter horizon, and field configurations which have support
around the horizon.

The general principles of holography and quantum gravity tell us that, at least in the limit $H \to 0$ where curvatures are low,
we expect the entropy to be associated to degrees of freedom at the
horizon of de Sitter space, located at a radius $H^{-1}$ from the observer.
The horizon represents the points which will (just) communicate
with the observer at the origin, with infinitely redshifted signals reaching him at future infinity. Every point within the horizon exists in the past lightcone of the asymptotic object `de Sitter space' whose entropy we want to compute.
\begin{figure}
\begin{center}
\epsfig{file=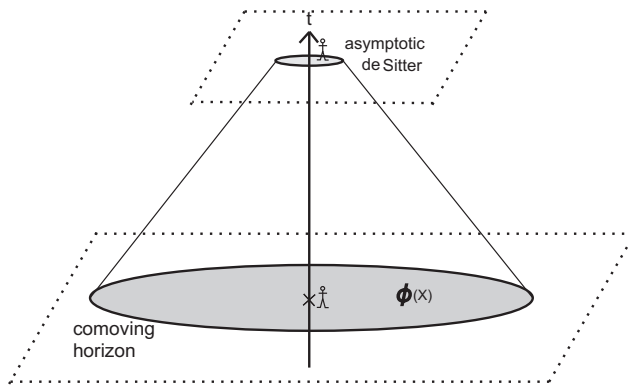,height=5cm}
\caption{The shaded area shows the comoving horizon together with an observer at the origin and the initial field profile $\phi(x)$. The future evolution of the shown observer leads to asymptotic de Sitter space. We associate the entropy of this quantum object `asymptotic de Sitter space' to the number of initial field configurations $\phi(x)$ that evolve in the far future to this object.}
\end{center}
\end{figure}

The association of the microstates with the horizon has heuristic support from consideration of the flat space limit.
Consider a flat scalar field theory in Minkowski space, with periodicity $f_a$. The axion is a canonical example of such a field theory.
By considering for example $\mc{N}=2$ compactifications of type II string theory on a Calabi-Yau it is clear that such field theories
do exist within quantum gravity.
Field theory in Minkowski space has an infinite number of `microstates'.\footnote{I thank John March-Russell for emphasising this and the following to me.} These are the superselection sectors associated with the
asymptotic value of the axion at infinity. These are defined by the map $S_{\infty}^2 \to \theta \in S^1$, where the asymptotic value of the axion field is $\theta$. These vacua are labelled by the continuous parameter $\theta$ and are infinite in number. They are superselection sectors as it would
cost an infinite amount of energy to transition from one sector to the other (i.e. to change the asymptotics of the scalar field).
As we move from Minkowski space to de Sitter space the sphere at infinity comes in to become the
de Sitter horizon and the infinite number of superselection sectors turns into a finite number of microstates. By (approximate) continuity in $H$ we expect these microstates to be associated with the allowed horizon configurations of $\phi$.

With a finite horizon there are two modifications to be made from the flat case.
First, a finite horizon means there is no longer an infinite energy cost to change the asymptotic profile of $\phi$. The sectors are no longer superselection sectors but are now connected by finite energy transitions. In particular it is appropriate to consider non-constant
profiles for $\phi$ on the horizon. Secondly, in either the flat space limit or the classical limit all possible values for $\phi$ represent distinct
states of the system. The states $\phi: S^2_{\infty} \to \phi_0$ and $\phi : S^2_{\infty} \to \phi_0 + \epsilon$ are distinct superselection sectors for arbitrarily infinitesimal $\epsilon$. This is no longer the case for the maps $\phi: S^2_{r=H^{-1}} \to \phi_0$ and
$\phi: S^2_{r=H^{-1}} \to \phi_0 + \epsilon$. de Sitter space has an irreducible finite temperature $T_{ds} = \frac{H}{2 \pi}$, and if $\epsilon$ is sufficiently small then these two maps should be regarded as the same quantum state, as the energy cost to move between them is much smaller
than the thermal energy of de Sitter space.

Having given a prescription for the entropy as the number of initial field configurations that asymptote to future de Sitter, we now try and count them. We focus first on bulk configurations that have no effect on scales beyond the horizon, and subsequently on horizon configurations.

\subsection{Bulk Microstates}

We first count microstates associated to internal degrees of freedom.
By internal degrees of freedom we mean field configurations
or profiles that are supported within the interior of the horizon and vanish asymptotically.
For distance scales much smaller than the Hubble scale, spacetime can be treated as flat and we use the standard formalism of flat space quantum field theory.
A localised particle-like wavepacket
is an example of such a configuration. All the energy of such a wavepacket is
localised in a finite region, and in particular the entire energy of the
quantum excitation lies within the horizon.

We want to count quantum states with no support outside the horizon. In effect, we work in a box with sides of size $H^{-1}$.
To ensure we asymptote to future de Sitter, we also
require that the energy density $\rho$ does not exceed a maximum value $\Delta V$. In practice it is easier to
 work with the corresponding bound on the total energy within the horizon,
$$
E_{total} < (H^{-1})^3 \ti \Delta V = H \left( \frac{\Delta V}{H^4} \right).
$$
It is easiest to count states in a
momentum space formalism. Bulk configurations have no support outside the horizon, are localised within a radius
$r \lesssim H^{-1}$ and thus all have energy $E \gtrsim H$. Furthermore, for energies much greater than $H$ we can treat the geometry as well approximated by Minkowski space and can use the formalism of QFT in flat spacetime. As a good approximation to our problem
we then count states of quantum field theory such that
\begin{enumerate}
\item
The energies of each mode are quantised in units of $H$ (as we work in a box of size $H$ and we consider modes with zero support outside this box).
\item
There are three spatial dimensions.
\item
The maximum total energy is $H \left( \frac{\Delta V}{H^4} \right)$.
\end{enumerate}
The counting of modes takes place as in free field theory. We will assume $N$ distinguishable massless bosons present.
For each dimension, each species has oscillators of energy $H, 2H, 3H, \ldots$ and each oscillator can have
an arbitrarily high occupation number.
The partition function for each species is
\bea
Z_{species} & = &
\left(1 + q^H + q^{2H} + \ldots \right)^3 \left( 1 + q^{2H} + q^{4H} + \ldots \right)^2 \left( 1 + q^{3H} + q^{6H} + \ldots \right)^3 \ldots \nn \\
& = & \frac{1}{(1-q^H)^3 (1-q^{2H})^3 \ldots }.
\eea
This partition function is esssentially that of the $\eta$ function,
$$
\eta(q) = q^{-1/8} \prod_{i=1}^{\infty} \frac{1}{(1 - q^n)}.
$$
We can neglect the $q^{-1/8}$ coefficient and for practical purposes we have
\be
Z_{species} = \eta(q^H)^{-3}.
\ee
For $N$ massless species, we have
\be
\label{yod}
Z_{overall} = \eta(q^H)^{-3N}.
\ee
The counting is more complex when there are massive modes with masses $H < m < H \left( \frac{\Delta V}{H^4} \right)$. As the presence of
masses
suppresses the allowed number of microstates, in this case equation (\ref{yod})
provides an upper bound on the bulk contribution to the entropy.

 We require the coefficient $C$ of the term in $\eta^{-3N}$ of order $q^{\Delta V / H^4}$.
 The total number of states with energies $E < E_{max}$ is then $\mc{N}_{total} < \left(\frac{\Delta V}{H^4} \right) C$.
 As $C$ turns out to be exponential in $\left(\frac{\Delta V}{H^4} \right)^{\half}$, the multiplicative prefactor gives only a logarithmic
 correction to the bulk entropy that can be neglected for large $\Delta V / H^4$.

The counting of asymptotic degeneracies of the $\eta$ function is a familiar problem from perturbative string theory (see for example
p117-8 of \cite{GSW}).
For the expansion of
$$
\left( \prod_{n} (1 - q^n) \right)^{-N} = \sum d_n q^n,
$$
it can be verified that the dominant (exponential) term in $d_n$ is
\be
\label{abc}
d_n \sim \exp \left( 2 \pi \sqrt{n} \sqrt{\frac{N}{6}} \right).
\ee
There is an additional power-law $n$ dependence which when considering the entropy is subdominant to the exponential term.
Eq (\ref{abc}) then gives an estimate of the bulk entropy as
$$
S_{bulk} \sim 2 \pi \sqrt{n} \sqrt{\frac{N}{6}}.
$$
As we count states up to a
maximum energy of $H \left( \frac{\Delta V}{H^4} \right)$,
we take $n \sim \frac{\Delta V}{H^4}$.

The contribution to the entropy from such internal modes is then
\be
S_{bulk} \sim \frac{2 \pi \sqrt{N}}{\sqrt{6}} \frac{\sqrt{\Delta V}}{H^2}.
\ee
For the `generic' case $\Delta V \sim V$, this gives a contribution to the entropy
\be
S_{bulk} \sim \sqrt{N} \frac{M_P}{H} \ll S_{dS},
\ee
provided that $\frac{H}{M_P}$ is small.

We note that if we were to allow the barrier height $\Delta V$ - effectively the maximum permissible energy density - to reach $M_P^4$, then the bulk entropy can
in principle saturate the de Sitter bound.
However we also note that in string-derived models, the maximum sensible barrier height is $\Delta V \sim m_s^4$ (rather than $M_P^4$).
This then gives
\be
S_{bulk} \sim \sqrt{N} \frac{M_s^2}{H^2}.
\ee
As in string models with low string scale the number of light degrees of freedom satisfies $N \lesssim \frac{M_P^2}{M_s^2}$, this gives
\be
S_{bulk} \lesssim \frac{M_P M_s}{H^2}.
\ee
This is always much less than the de Sitter entropy $S_{dS}$ unless $M_s \sim M_P$ and the barrier height is again at the Planck scale.

From the above, the only way bulk entropy can make up a significant fraction of the de Sitter entropy is \emph{either} if we allow
Planck scale energy densities without triggering destabilisation, \emph{or} if our asymptotic de Sitter future has $\frac{H}{M_P}$ sufficiently large
that terms of order $\frac{M_P}{H}$ cannot be neglected compared to terms of order $\frac{M_P^2}{H^2}$.
At least in string derived models the former cannot arise, whereas we implicitly exclude the latter case by requiring that we can sensibly talk about a classical inflating geometry (in any case, for inflationary models $\frac{H}{M_P} \lesssim 10^{-4}$ and so this approximation should be good).

It is worthwhile here to restate an important if obvious point. When $H \ll M_P$ (for example when $\frac{M_P^2}{H^2} \sim 10^{120}$ as in our world),
the number of de Sitter microstates is truly gargantuan ($10^{10^{120}}$ for our world). While a bulk entropy of $S_{bulk} \sim \frac{M_P}{H}$ appears superficially similar to $\frac{M_P^2}{H^2}$, in terms of microstates it gives essentially no contribution: for all practical purposes we are still missing \emph{all} the microstates.

This tells us that field configurations and excitations which are localised entirely within the horizon have a negligible contribution
to the de Sitter entropy. This is consistent with expectation, as we expect the entropy of a gravitational horizon to be associated with
a lack of knowledge of degrees of freedom beyond that horizon. For de Sitter, this must correspond to profiles having support outside the de Sitter horizon, which the internal bulk configurations considered above most certainly do not.

\subsection{Horizon Entropy}

We next consider entropy associated to degrees of freedom not localised in the interior. These approximately correspond to
field configurations that modify the horizon vevs. The construction of such vevs requires quantum states which have support both in and out of the horizon. This is easiest to see by thinking about short-distance field variations, when the distance associated to the variation
is much smaller than the horizon scale and we can use flat space QFT.

The elementary excitation of a quantum field is the infinitely delocalised
1-particle state with energy $E$ and momentum $k$. By constructing a wavepacket, we can localise such an excitation over a finite region of space time $L^3$. The local energy density of this state is $E/L^3$ and the energy $E$. Physically, there is nothing wrong with a delocalised excitation where
the delocalised region is part within the de Sitter horizon and part beyond. Provided the delocalised region all fits within some horizon,
there exists an observer for whom  this is just a
 conventional wavepacket. Furthermore, the horizon has no intrinsic existence that can be measured locally: it
 has only been defined relative to the distant observer it
can communicate with in the far future. This is illustrated in figure 3.
\begin{figure}
\begin{center}
\epsfig{file=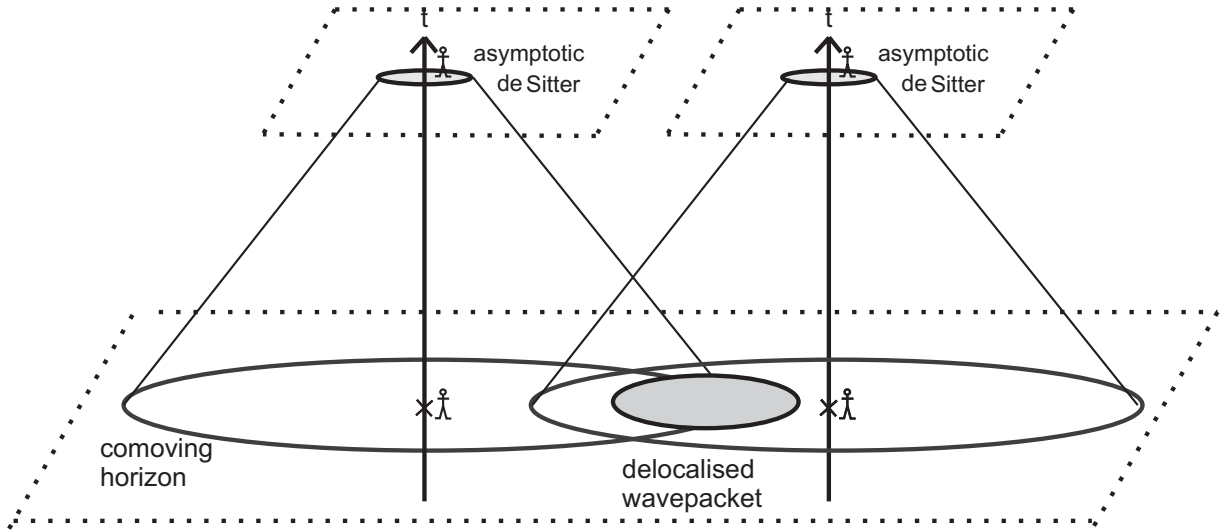,height=7cm}
\caption{A wavepacket delocalised across the horizon. From the viewpoint of an observer whose horizon encompasses this wavepacket, this is just
a conventional delocalised quantum mechanical state.}
\end{center}
\end{figure}

From the viewpoint of our original observer these configurations are part within the horizon and part beyond.
This explains why the bulk counting of states does not apply. First, such configurations change the boundary conditions on the field vev, and
secondly, as they are delocalised outside the horizon
the energy they contribute within the de Sitter horizon can be much less than the 1-particle excitation energy $E$. There is not
a `minimum energy quantum' of $H$ as there was previously for the bulk entropy.

We now provide two ways of calculating the boundary entropy associated to such states.
The first comes from an attempt to count microstates explicitly, and the second from
considerations of entanglement entropy.

On the first approach, we regard the continuum theory of a scalar field as a long-distance approximation to a more fundamental
latticised model. We discretise a field $\phi$ from a continuum variable $\phi(x)$ to a set of discrete values $\phi_i(x)$ where $i$ labels
the lattice point. The points $i$ and $i+1$ are separated  by the lattice spacing $a$.
We can continue discretising $\phi$ to shorter and shorter lattice spacings $a$ until we reach
a fundamental scale $a_{min}$ beyond which we are not permitted to discretise any further. $a_{min}^{-1}$ then plays the role of the
ultraviolet energy cutoff for this field. (For the case of axions, we will relate $a_{min}$ to $f_a$ below).

On this picture, the existence of an ultraviolet cutoff $a_{min}^{-1}$ allows us to discretise the field $\phi$ down to a lattice
with nodes separated by $a_{min}$. At this discretisation scale, we have obtained the minimum box size at which it is sensible to talk about
independent choices of states for $\phi_i$ and $\phi_{i+1}$. In this limit, we can count two possible states of $\phi$ per lattice point, and the
state of $\phi$ at point $i$ does not determine the state of $\phi$ at point $i+1$. This is the lattice version of `choosing the field configuration at time zero'.

Recall that our working definition of the de Sitter entropy is the number of initial field configurations that asymptote to future de Sitter.
As the horizon field profiles only communicate with future de Sitter in the asymptotically redshifted future, all allowed profiles contribute to the
entropy independent of their original horizon energy density (which can satisfy  $\rho \gg \Delta V$). The structure of these profiles is illustrated in figure 4.
\begin{figure}
\begin{center}
\epsfig{file=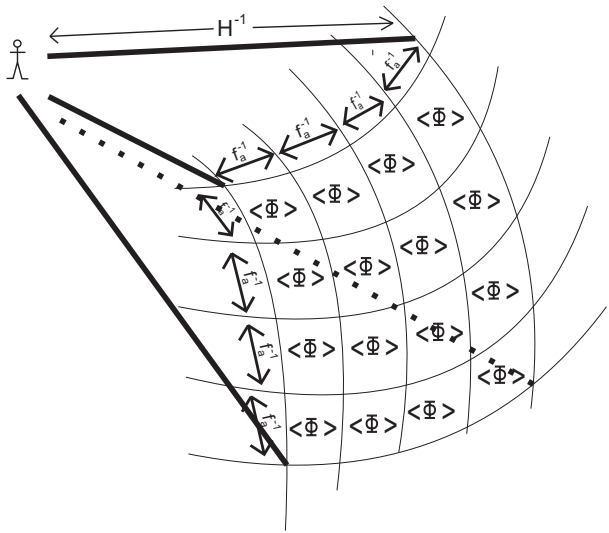,height=7cm}
\caption{The horizon field configurations, set by the vev of $\Phi$ in each fundamental box. This is shown for an axion field where the fundamental
box size is $f_a^{-1}$.}
\end{center}
\end{figure}
For a minimal box size of $a_{min}$,
there are then a total of $2^{\frac{4 \pi H^{-2}}{a_{min}^2}}$ horizon microstates of the field $\phi$, giving  a contribution to the entropy
of
\be
\label{gfd}
S_{horizon} = 4 \pi \ln 2 \frac{a_{min}^{-2}}{H^2},
\ee
associated to the different horizon profiles of $\phi$ contributing to asymptotic future de Sitter.

Note that this estimate has involved resolving the continuum field $\phi$ down to the ultraviolet scale $a_{min}$, and examining the states
of $\phi$ at this scale, where $a_{min}$ is effectively the smallest scale at which the field $\phi$ exists as a useful quantity.
In particular, the estimate (\ref{gfd}) involves the physics of the field $\phi$ only at the energy scale
$a_{min}^{-1}$. As mass parameters are relevant in the infrared and irrelevant in the ultraviolet, this also implies that the above estimate will
also hold for any field having a mass $M < a_{min}^{-1}$: such a mass is irrelevant at the scale $a_{min}^{-1}$.

For a model with many species $N$
lighter than the ultraviolet cutoff, the total number of states would have an additional factor of $N$ in the exponent and
the above estimate of the total boundary entropy should then be multiplied by $N$, giving
\be
\label{Ngfd}
S_{horizon} = 4 \pi \ln 2 \frac{N a_{min}^{-2}}{H^2}.
\ee

We discuss the relation of $a_{min}$ to $f_a$ in section 4 below, but first give
a second estimate for the horizon entropy of a field with periodicity $f_a$.

We have observed that the construction of field theory states with a vev for $\phi$ on the boundary necessarily requires wavepackets which are delocalised across the horizon.
From the viewpoint of the observer at the boundary, this is a conventional entangled state with all parts of the wavefunction accessible to observation. However by definition the observer at the origin can never access any of the information contained outside the horizon and can never know about the wavefunction outside the horizon.

This suggests another way of thinking about the boundary entropy: as the entanglement entropy of the scalar field across the horizon.
Entanglement entropy is defined as follows. Consider the ground state density matrix, $ |0><0|$.
The ground state can be written as
$$
|0>  = \sum_{ij} c_{ij} |i>_{in} |j>_{out}.
$$
Tracing the density matrix over all external (or internal) states, we obtain a new density matrix,
$$
\rho_{new}  = \sum_j <j|_{out} \left( |0><0| \right)  |j>_{out} = \sum M_{ik} |i>_{in} <k|_{in}
$$
The entanglement entropy is the entropy $S = -\sum_i p \ln p_i$ of the eigenvalues of the density matrix $M$.

Entanglement entropy has been studied in many works with seminal papers including \cite{sorkin, srednicki}.
Useful reviews are \cite{11043712, casini}.
For our purposes the most important result is that
the dominant contribution to entanglement entropy depends on the area of the surface bounding the two regions of space and the ultraviolet cutoff, behaving as
\be
\label{ent}
S_{entanglement} = \lambda M^2 A
\ee
where $M$ is the ultraviolet cutoff $M = a^{-1}$ for a lattice regularisation, $A$ is the area of the horizon and $\lambda$ is a proportionality constant. There are additional constant/logarithmic `universal' terms which do not depend on the ultraviolet cutoff.
For many applications these universal terms are the most interesting, as they are insensitive to the UV and so can track
changes in the low energy behaviour.
However as the ultraviolet cutoff
is key to the physics we are considering we focus only on the leading term.

The significance of (\ref{ent}) is that entanglement entropy is both local and UV-sensitive (it depends on the UV cutoff and the bounding area).
The entanglement entropy can be defined in non-gravitational theories and the horizon that enters in it does not need any intrinsic significance. This makes entanglement entropy an appropriate quantity for de Sitter space where the horizon is observer-dependent and requires a choice of origin for its definition.

The entanglement entropy for a scalar field across the de Sitter horizon is given by
$$
S_{entanglement, axion} = 4 \pi \lambda \frac{a_{min}^{-2}}{H^2},
$$
for some $\lambda$. This has the same parametric scaling (and the same dependence only on the UV structure) as our previous estimate in terms
of horizon degrees of freedom. As before, the precise mass of the field is irrelevant: for masses $m \ll a_{min}^{-1}$
the entanglement entropy of a massive field is equivalent to that of a massless field.

It is not so easy however to determine the constant $\lambda$. Indeed as the result is UV sensitive the determination of $\lambda$ depends on
precisely how the theory is regularised.
For flat Minkowski space, there exists the  classic calculation of Srednicki \cite{srednicki}.
This involves a specific radial regularisation. The field $\phi$ is decomposed into 3d spherical harmonics, and
 space is split into continuum shells with a radial coordinate quantised in units of a fundamental scale $a_{min}$.
The entanglement entropy can then be calculated numerically, giving
$$
\lambda \sim \left( \frac{0.30}{4 \pi} \right).
$$
There are two questions as to the applicability of this computation to our case.
First, the radial regularisation is one specific choice which involves a discretisation purely in the
 radial direction. This is not obviously the correct regularisation for an axion field with well-defined periodicity.
Secondly, there is the question of whether Minkowski calculation can be carried over to a weakly de Sitter case (in favour: entanglement entropy is UV sensitive; against: the area that appears involves a shell of characteristic size equal to the Hubble radius).

Setting aside these caveats the Srednicki regularisationgives
$$
S_{entanglement} \simeq 0.3 \frac{a_{min}^{-2}}{H^2}.
$$
This differs from the earlier `bits per unit area' approach by a factor of 20.

Up to the numerical factors both calculations give similar parametric scalings and have a similar dependence only on the UV structure (the entanglement
entropy of a massive field is identical to a massless one provided the mass of the field is much smaller than the UV cutoff). As before, this is because the dominant contribution to entanglement entropy comes from short-distance fluctuations across the horizon. Mass terms are IR-relevant
and so, provided the mass is much less than the ultraviolet cutoff, represent a long-distance effect.

An axion field comes with an intrinsic scale $f_a$. This scale is associated to both the field periodicity and the mass and tension
of excited axion strings. For an axion field the parametric scaling of the ultraviolet cutoff is clearly $\mc{O}(f_a)$. Let us now try and make this estimate more precise.

\section{What is the Ultraviolet Cutoff?}

We suppose we have a field with periodic field range $\phi \equiv \phi + f_a$. What is the ultraviolet cutoff on such a field?

A periodic field contains an intrinsic ultraviolet cutoff. For models where the field arises from
spontaneously breakdown of a global symmetry (for example as a Goldstone boson) this cutoff is associated to the energy scale
at which the symmetry is restored. At this scale, the concept of an axion vev ceases to be meaningful. A good analogy is the case of the
pion, which similarly arises as a Goldstone boson. The pion vev ceases to be a meaningful concept at energies $E \gg \Lambda_{QCD}$, where chiral
symmetry is restored and the pion no longer exists as a well-defined particle.
From a low-energy perspective the axion UV cutoff is associated to the presence of new states, such as axion strings, whose
existence depends on the field periodicity.

Both the above approaches to calculating entropy involved the discretisation of a continuum theory. In both cases the ultraviolet
cutoff scale was identified as the scale $a_{min}$ of the minimal discretisation. Let us adopt this process for an axion model and
see when we cease to get meaningful structure.
We replace the continuum coordinate $x$ with a discrete set of points $x_n$, where $x_n = n a$ and $a$ is the lattice spacing, and the continuum
field $\phi(x)$ with the discrete values $\phi(x_n)$.

The introduction of a lattice of period $a$ is equivalent to a UV cutoff on momentum modes, $p \leq \frac{2 \pi}{a}$. There is also an infrared cutoff
associated to a finite size lattice, but this will not be important to our purposes. Recall that for no UV cutoff
the free field 2-point function is
\be
\langle \phi(x) \phi(y) = \int \frac{d^3 p}{(2 \pi)^3} \frac{1}{2 E_p} e^{-ip \cdot (x -y)} \to \frac{1}{4 \pi^2 (x-y)^2} \hbox{ for no UV cutoff}.
\ee
The variance of the field at any given point is
\be
\langle \phi(0) \phi(0) \rangle = \frac{1}{4 \pi^2} \int d \vert p \vert \vert p \vert = \frac{p_{max}^2}{8 \pi^2}.
\ee
Equating $p_{max}$ with $2 \pi / a$, we thereby obtain $\langle \phi(0) \phi(0) \rangle = \frac{1}{2 a^2}$.

Discretisation replaces a continuum field taking values at each individual point with a lattice
where the field is restricted to individual points. This process carries the implicit assumption that the expression
`vev of $\phi$ at a point' is meaningful: i.e. that there is more than one state that can be associated to $\phi$ at each point. This
provides a limit to the discretisation: once $\langle \phi^2(0) \rangle \gtrsim f_a^2$, the wavefunction for $\phi(0)$ is necessarily distributed over all possible field values, and we can no longer give meaning to $\langle \phi(0) \rangle$: quantum scatter unavoidably
diffuses $\phi$ over the entire field range.
We therefore determine the `minimum discretisation' by
$$
\langle \phi(0) \phi(0) \rangle = \frac{1}{2 a_{min}^2} = f_a^2,
$$
and so we take $a_{min} = \frac{1}{f_a \sqrt{2}}$. This gives an appropriate estimate of the scale $a_{min}$ we can use
when discretising the axion field.

We can now make our two approaches of section 3 more quantitative.
Using $a_{min} = \frac{1}{f_a \sqrt{2}}$, the `bits per unit area' approach gives
a total of $2^{\frac{8 \pi f_a^2}{H^2}}$ horizon microstates of the field $\phi$, giving a contribution to the entropy
of
\be
S_{horizon} = 8 \pi \ln 2 \frac{f_a^2}{H^2} \simeq 17 \frac{f_a^2}{H^2},
\ee
associated to the different horizon profiles of $\phi$ contributing to asymptotic future de Sitter.
Using entanglement entropy and the Srednicki regularisation gives
$$
S_{entanglement} \simeq 0.6 \frac{f_a^2}{H^2}.
$$
This differs from the earlier `bits per unit area' approach by a factor of 20.

Both of these are to be compared with the de Sitter entropy $S_{dS} = \frac{8 \pi^2 M_P^2}{H^2}$.
We can now use these to give a bound on the allowed number of axions in de Sitter space with decay constant $f_a$. This bound comes from
requiring that the calculable entropy associated to the axions is smaller than the de Sitter entropy. Depending on the method employed, we obtain
a bound
$$
\sqrt{N} f_a < \sqrt{\frac{\pi}{\ln 2}} M_P \qquad \hbox{(bits per area)}
$$
or
$$
\sqrt{N} f_a \lesssim \pi \sqrt{13} M_P \qquad \hbox{(entanglement entropy)}
$$
This makes it clear that parametrically transPlanckian flat directions cannot be obtained by the combination of many flat
subPlanckian directions: the attempt will violate fundamental principles of quantum gravity.

The above results give the parametric scaling of the entropy with field range and number of species.
However they are not a rigorous determination of the numerical prefactor (and indeed the two approaches give prefactors that differ by
a non-trivial amount). It would be useful to develop a precise method to compute these prefactors.
In this respect, string theory provides an appropriate arena for direct calculation, as it gives a explicit
UV regularisation. We leave this more precise study of the UV cutoff for future work \cite{JCLukas}.

We also comment that
in the context of effective theories derived from string theory, the constraints on such models only become more severe.
As we have seen, the entanglement entropy is
an ultraviolet sensitive quantity that depends on the length scale of the ultraviolet cutoff. Therefore
\emph{any} particle with a mass significantly lighter than the cutoff counts as effectively massless when computing entanglement entropy.
This implies that in addition to the above contributions to entanglement entropy from
the axions, there will also be contributions
from \emph{all} other fields with masses below the string scale. We note this also includes the graviton, which is assuredly a light
degree of freedom.

Even in the most minimal case (provided the supersymmetry breaking scale is lower than the string scale), this will include the saxions, the supersymmetric partners of the axions, and the
 fermionic partner the axino. Fermionic degrees of freedom contribute to entanglement entropy with a factor of one half
 that of bosonic modes \cite{9403108},
 and so the above calculable contributions to entanglement entropy are immediately tripled.
 For geometric compactifications, Kaluza-Klein replicas of the axions and partners will
 also contribute, further increasing the entropy.
We discuss this further in section 6 but leave a full discussion to the future \cite{JCLukas}. We first however
discuss other inflationary models.

\section{Chaotic Inflation}

N-flation was an attempt to obtain trans-Planckian field ranges through the combination of many fields with controlled sub-Planckian
field ranges. A more traditional field-theoretic approach to obtaining trans-Planckian field ranges is simply to write them down.
Examples are models of chaotic inflation, based on for example $m^2 \phi^2$ or $\lambda \phi^4$ potentials.
In these cases there is a single field with a transPlanckian field range with a potential that is approximately flat over
a transPlanckian distance. These theories are described by a Lagrangian
\be
\label{inflate}
S = \frac{M_P^2}{2} \int d^4 x \sqrt{G} \left( R - \half \partial_{\mu} \phi \partial^{\mu} \phi - V(\phi) \right),
\ee
where $V(\phi) = \half m^2 \phi^2$ or $V(\phi) = \frac{1}{4} \lambda \phi^4$. The argument for the consistency of such models
is that as long as $V(\phi) \ll M_P^4$ there is no breakdown in the effective field theory. The counter-argument is that such models
contain no reason to exclude from the potential general terms of the form $M_P^4 \lambda_n \left( \frac{\phi}{M_P} \right)^n$, which
for $\langle \phi \rangle > M_P$ destroy slow-roll.

Let us apply our above approach to such models.
One can always view an unbounded scalar field as in (\ref{inflate})
as having infinite periodicity: i.e. as the limit of an axion field of periodicity $f_a$
in the limit that $f_a \to \infty$. The potential $V(\phi)$ in (\ref{inflate}) can then be regarded as being periodic on a much larger scale
than is relevant for the inflationary epoch. We can ask what entropy is associated to the scalar field in (\ref{inflate}). As described above,
this is determined by the length scale down to which we can discretise the field $\phi$. No cutoff scale appears in (\ref{inflate}) earlier than
the Planck scale. Furthermore, the fact that the potential is (at least to a good approximation) flat on super-Planckian distances tells
us that this does not obstruct a discretisation down to scales $a_{min} \sim M_P^{-1}$: discretising to a scale $a_{min}$ sets the variance of the
field as
$$
\langle \phi(0) \phi(0) \rangle = \frac{1}{2 a_{min}^2}.
$$
Note that this requires the flatness of the potential on the scale $a_{min}^{-1}$, as this expression originates
as the free field expression
$$
\langle \phi(0) \phi(0) \rangle = \int \frac{d^3 p}{(2 \pi)^3} \frac{1}{2\sqrt{p^2 + m^2}}.
$$
The presence of
large interaction terms (or the deviation from approximate flatness across a range $\Delta \phi \sim a_{min}^{-1}$)
would destroy the validity of this free field calculation.

The discretisation of the model (\ref{inflate}) can then be carried out without problem down to length scales comparable to the Planck scale,
giving an entropy for the scalar field
\be
\label{zzz}
S = \lambda^{'} \frac{M_P^2}{H^2},
\ee
as the ultraviolet cutoff on the model is around the Planck scale from the non-renormalisable terms in the Einstein-Hilbert action.
There is also a similar contribution from the graviton degree of freedom for which the UV cutoff is also around the Planck scale.

As chaotic inflation models are not UV complete, we cannot
say any more about $\lambda^{'}$. However as (\ref{zzz}) approximately saturates the de Sitter entropy bound, we can say that such models
are sitting on an incalculable fence separating viable and non-viable theories.
This identifies an intrinsic problem with chaotic inflation models that goes beyond the simple question of how to realise them or how to ensure
the vanishing of coefficients non-renormalisable operators: there is no guarantee of the consistency of such models
with quantum gravity.

\section{Inflationary and de Sitter String Constructions}

For inflationary or de Sitter models that are meant to come from string theory, we can go further.
For a string model, the ultraviolet cutoff is not a free parameter but is set by the model (as the moduli vevs determine the ratio of the
string and Planck scales).
This allows us to exclude or set constraints on certain string-inspired models that have nothing obviously wrong with them other than looking ugly.
One example is racetrack inflation \cite{0406230}, which effectively involves two condensing gauge groups $SU(100) \ti SU(90)$.
An $SU(100)$ gauge group has at least $10^4$ gluons and gauginos associated to it (there may also be additional KK replicas which we do not
count). For consistency, the condensation scale of the gauge group has to be much less than the ultraviolet cutoff. Then, while the gluons and
gauginos are heavy compared to the inflationary scale, they are light compared to the ultraviolet cutoff and count towards the de Sitter entropy as if they were massless particles. The analysis of section 3 therefore carries through with $\gtrsim 10^4$ light species.

This number of light species implies the UV cutoff (i.e. the string scale)
must be correspondingly reduced, and no bigger than $M_P/\sqrt{10^4}$. However inspection of the proposed parameters of the models reveals that the string scale is actually much higher
than this,\footnote{This follows from the fact that in the model
$\frac{R^4}{g_s} = \hbox{Re}(T)  \simeq 120$ in natural units and so the string scale is close to the Planck scale.}  and so as a
string model this is internally inconsistent.

A similar critique extends to the KKLT construction of de Sitter vacua \cite{0301240}. The volume is stabilised via a non-perturbative
superpotential
$$
W = W_0 + A e^{-aT},
$$
with $a = \frac{2 \pi}{N_c}$ for gaugino condensation of an $SU(N_c)$ gauge group. The stabilisation sets
 $e^{- a T} \sim W_0$, and so $\frac{R^4}{g_s} = \hbox{Re}(T) = \frac{N_c}{2 \pi} \ln W_0$.
Large $N_c$ is then necessary to obtain stabilisation in the geometric regime ($R > 1$).
However if the volume-stabilising term is generated by gaugino condensation of an
$SU(N_c)$ gauge group, then there are at least $(N_c^2 - 1)$ light gluons, with masses below the UV cutoff
and contributing to the entropy.

Therefore while large $N_c$ appears beneficial in obtaining control, it actually does not help:
any gain obtained by taking $R^4 \to N_c R^4$ by having $N_c \gg 1$
is offset by a corresponding decrease in the maximal allowed UV cutoff $\Lambda_{UV} \to \frac{\Lambda_{UV}}{N_c}$.
For the parameters in the original paper \cite{0301240}, $a=0.1$ implying $N_c = 60$ and at least $\gtrsim 3600$
degrees of freedom below the UV cutoff. The entropy considerations in this paper then require the string scale to satisfy
$$
m_s \lesssim \frac{M_P}{60}.
$$
By examining the moduli stabilisation parameters and using
the standard relation $m_s = \frac{g_s M_P}{\sqrt{\mc{V}}}$ in natural units, it is clear that this is not
satisfied and again the model is internally inconsistent.

These problems arise from the combination of a large numbers of degrees of freedom together with a UV cutoff (i.e. string scale)
close to the Planck scale. If the UV cutoff is much lower than the Planck scale, then the presence of large numbers
of degrees of freedom ceases to be a problem. This situation is realised in the
LARGE Volume Scenario, which has the big advantage of a
large parametric separation of scales that enables one to make an analysis of the origin of de Sitter entropy much cleaner.

To recall: the LVS stabilises the bulk volume at exponentially large volumes in a Swiss-cheese type geometry.
This involves one large cycles and several small blow-up cycles. As the volume is exponentially large, the string scale
is exponentially small (compared to the Planck scale). The original minimum is AdS, with a depth of $V_{AdS} \sim - m_{3/2}^3 M_P$.
This minimum is then uplifted to de Sitter.

The key point is that the structure of the minimum allows the vacuum to be realised at large volume, weak coupling and dilute fluxes.
The worldsheet theory is then weakly coupled in both the $\alpha'$ and $g_s$ expansion. This implies that the counting of worldsheet modes
and degrees of freedom can be approximated by those of the free theory.

We leave for future work \cite{JCLukas} a full analysis. However we can already see parametrically where the de Sitter entropy can
arise from microscopically. At large volumes, the number of degrees of freedom lighter than the string scale is set by the bulk KK modes.
As $R_{bulk} \to \infty$, the number of such KK modes grows as $R_{bulk}^6$: each mode is quantised in $R_{bulk}^{-1}$ and
so these modes approximately occupy a six-dimensional sphere with unit mode density and radius $R_{bulk}$.
The advantage of the limit $R_{bulk} \gg 1$ is that the contributions from bulk zero modes,
brane KK modes, KK modes localised on small cycles, etc all are subleading in this limit, as they scale with no or smaller powers of $R_{bulk}$.
For example, the number of KK modes localised on a bulk 7-brane scales as $R_{bulk}^4$ and the number of chiral matter zero modes
scales as $R_{bulk}^0$. The limit $R_{bulk} \gg 1$ (which, importantly, is an allowed limit in LVS) therefore isolates the bulk
closed string KK modes as the numerically dominant fields and decouples the more model-dependent modes from the counting.

We have established above that such bulk modes will give a contribution to the de Sitter entropy of
\be
\label{cow}
S \sim \mc{N} \frac{\Lambda_{UV}^2}{H^2}.
\ee
As $\mc{N} \sim R^6$, and $\Lambda_{UV} \sim \frac{M_P}{\sqrt{V}}$, we see that in the limit $R \to \infty$, we obtain
$$
S \sim \frac{M_P^2}{H^2}.
$$
This has the correct parametric scaling to give the de Sitter entropy. The advantage of the large-volume limit is that it has isolated
the universal features of the compactification while rendering the non-universal features (e.g.
the details of branes/orientfiolds and how many blow-up moduli are present) subleading.
In black hole physics a similar formula and approach to (\ref{cow}) has been discussed in \cite{08063976, 08121940}.
We will investigate (\ref{cow}) further
in the future \cite{JCLukas}.

\section{Conclusions and General Comments}

Inflation is a promising candidate theory of the early universe built around the formalism of qauntum field theory in
curved spacetime. The approach of this paper has been to consider bounds on inflation that originate from
quantum gravity. We have done so by requiring that the entropy associated to the fields necessary to build
the inflationary model be smaller than that of the de Sitter entropy. Such an approach has teeth and in particular
constrains attempts at obtaining transPlanckian flat directions: the entropy saturates the de Sitter bound at
approximately the same point at which transPlanckian vevs can be obtained.

This approach also constrains small-field models, when the inflation model is realised with a high UV cutoff
and a large number of degrees of freedom (the example we gave was racetrack inflation). Similar arguments
constrain the KKLT construction of de Sitter vacua (and other similar ones with large-rank gauge groups and a relatively
small stabilised volume).
The essential problem is the same:
if a model of de Sitter space has more microstates than allowed by quantum gravity in de Sitter space then it cannot be consistent.

We have mainly studied bounds on flat directions in exact de Sitter space. Inflation does not involve exact de Sitter but only an approximate de Sitter solution that is terminated at the end of inflation. For slow roll inflation the deviation from exact de Sitter is parametrised by the slow-roll
parameters $\epsilon, \eta \ll 1$ and we expect our results to hold up to corrections proportional to the slow roll parameters.
It would be interesting to make this last statement more precise.

We have used two approaches to determining the entropy: `bits per unit area' and entanglement entropy. These give the same
parametric scaling but differ by a numerical constant. It would be obviously desirable to refine these calculations and determine
the `right' answer. As the uncertainties closely relate to the UV cutoff, an explicit string calculation of the relevant effect would be desirable.

Finally, this work is ultimately
motivated by an observable: tensor modes in the CMB, that can be measured by the Planck satellite and correlate with transPlanckian field excursions.
We look forward to experimental guidance on this question.

\acknowledgments{I thank Mathew Bullimore, Matthew McCullough, Francisco Pedro, Steve Simon, Lukas Witkowski and particularly John March-Russell for discussions. I am funded by the Royal Society and New College, Oxford.}

\end{document}